\documentclass[smallextended]{svjour3}

\usepackage{amsmath,graphicx,hyperref}
\usepackage{booktabs}
\usepackage{array}
\usepackage{float}
\usepackage{pifont}
\usepackage[sort&compress,numbers]{natbib}
\usepackage[linesnumbered,ruled,vlined]{algorithm2e}
\usepackage{comment}
\usepackage{tabularx}
\usepackage{geometry}
\usepackage{graphicx}
\usepackage{longtable}

\begin{document}

\title{Predictive-CSM: Lightweight Fragment Security for 6LoWPAN IoT Networks
}


\author{Somayeh Sobati-Moghadam      
}


\institute{Somayeh Sobati-Moghadam *, ** 	
	\at 
		\(^{*}\) Hakim Sabzevari University, Sabzevar, Iran. Email: s.sobati@hsu.ac.ir  \\
	\(^{**}\) Department of Computer Engineering, Ferdowsi University of Mashhad, Mashhad, Iran
}       

\date{Received: date / Accepted: date}

\maketitle

			\begin{abstract}
				Fragmentation is a routine part of communication in 6LoWPAN-based IoT networks, designed to accommodate small frame sizes on constrained wireless links. However, this process introduces a critical vulnerability: fragments are typically stored and processed before their legitimacy is confirmed, allowing attackers to exploit this gap with minimal effort.
				
				In this work, we explore a defense strategy that takes a more adaptive, behavior-aware approach to this problem. Our system, called Predictive-CSM, introduces a combination of two lightweight mechanisms. The first tracks how each node behaves over time, rewarding consistent and successful interactions while quickly penalizing suspicious or failing patterns. The second checks the integrity of packet fragments using a chained hash, allowing incomplete or manipulated sequences to be caught early, before they can occupy memory or waste processing time.
				
				We put this system to the test using a set of targeted attack simulations, including early fragment injection, replayed headers, and flooding with fake data. Across all scenarios, Predictive-CSM preserved network delivery and maintained energy efficiency, even under pressure. Rather than relying on heavyweight cryptography or rigid filters, this approach allows constrained devices to adapt their defenses in real time—based on what they observe, not just what they’re told. In that way, it offers a step forward for securing fragmented communication in real-world IoT systems.
			\end{abstract}


\section{Introduction}

These days, low-power wireless networks are doing a lot of heavy lifting in the world of IoT. Whether it’s managing irrigation systems in agriculture, monitoring machinery in factories, or helping homes run more efficiently, these networks are everywhere. One technology that’s made this possible is 6LoWPAN. It allows tiny devices with minimal memory and power to speak IPv6—essentially giving them a seat at the table on the global internet, even if they’re running on coin cell batteries and a few kilobytes of RAM.

But as useful as 6LoWPAN is, it comes with a trade-off. Since these small devices can’t send big packets all at once, the data has to be broken into fragments. That sounds fine in theory, but in practice it creates a weak spot. Devices tend to accept these fragments as they arrive and hold them in memory, even before knowing if the whole message makes sense or comes from someone trustworthy. That small gap—between accepting and verifying—gives attackers just enough room to cause trouble.
 Since fragments are typically accepted and held in memory before the complete packet is reassembled and validated, attackers can exploit this gap. Even without cryptographic keys or full control of the network, a malicious node can cause serious disruption simply by manipulating how fragments are handled.This opens the door to a range of attacks that target the reassembly buffers of constrained devices. These include not only traditional buffer-reservation attacks but also more nuanced behaviors like clone flooding, header replay, and timing-offset injections—many of which have been observed in recent threat modeling studies~\cite{zhang2022review, ghubaish2021comprehensive}.

For example, in a simple buffer-reservation attack, an adversary sends incomplete packet fragments (often just the initial FRAG1, which refers to the first fragment of a 6LoWPAN packet that carries essential header information and initiates the reassembly process) slightly before a legitimate transmission. Since devices have very limited buffer space—typically only one or two slots—this alone is enough to block legitimate communication. More advanced strategies involve injecting full but meaningless fragment sequences to exhaust reassembly logic, replaying cloned headers to impersonate trusted nodes, or introducing fragment delays to confuse packet sequencing and induce timeouts. These low-cost, high-impact attacks can silently degrade network performance or cause complete denial of service.

Existing mitigation strategies offer only partial solutions. Stateless 6LoWPAN stacks, such as those implemented in Contiki-NG, have no native fragment validation and accept traffic purely based on header structure and timing. Trust-based systems like Chained Secure Mode (CSM)~\cite{rodrigues2020chained} attempt to mitigate this by filtering packets from untrusted routes, but they operate only at the routing layer and are blind to fragment-level anomalies. On the other end of the spectrum, cryptographic approaches such as SecuPAN~\cite{hossain2018secupan} enforce MAC-based validation of every fragment, but at a cost that is often too heavy for battery-powered or low-RAM devices.

In this paper, we present \textit{Predictive-CSM}, a practical and resource-aware defense framework for 6LoWPAN fragmentation attacks. The system combines two lightweight mechanisms: an adaptive trust model that continuously learns from a neighbor's fragment-level behavior, and a chained hash validator that verifies the integrity of each fragment incrementally. Unlike traditional models, Predictive-CSM operates directly at the adaptation layer, where fragmentation occurs, enabling the system to identify and discard suspicious fragments before they trigger memory exhaustion or communication breakdowns.

We evaluate Predictive-CSM across five carefully designed attack scenarios: early FRAG1 injection (buffer-reservation), full-fragment flooding, header-replay cloning, burst injection (high-rate FRAG1 spam), and late-phase injection (timing-offset spoofing). These scenarios are chosen to reflect both known attacks and realistic adversarial behavior patterns. Simulation results showed that our approach maintains high packet delivery success, rapid adversary detection, and low energy overhead under all attack conditions. By aligning security enforcement with the actual layer where fragmentation occurs, Predictive-CSM closes a longstanding gap in 6LoWPAN security. It offers a deployable, efficient, and adaptive solution that directly addresses the operational constraints of real-world IoT networks while resisting some of the most effective low-layer threats known to date.
\paragraph{Paper Organization.}
The remainder of this paper is structured as follows: Section~\ref{sec:related-work} reviews existing approaches to 6LoWPAN fragmentation security and highlights their limitations. Section~\ref{sec:proposed-method} introduces the architecture of the proposed Predictive-CSM framework, including the Predictive Trust Engine (PTE) and Fragment Signature Validator (FSV). Section~\ref{sec:protocol-design} presents the detailed protocol design, including message formats, trust evaluation logic, and fragment validation workflow. Section~\ref{sec:experimental-setup} describes the simulation environment, attack scenarios, and evaluation metrics used to assess performance. Section~\ref{sec:results} provides a comprehensive analysis of the experimental results under multiple adversarial conditions. Section~\ref{sec:analytical-model} develops an analytical model of trust dynamics, buffer utilization, and cryptographic validation. Finally, Section~\ref{sec:conclusion} concludes the paper and outlines directions for future work.

\section{Related Work}\label{sec:related-work}

		Recent advances in securing 6LoWPAN networks have led to various approaches addressing fragmentation vulnerabilities. For instance, Sharma et al. \cite{sharma2021adaptive} proposed an adaptive trust management system that leverages machine learning techniques to identify and mitigate attacks on IoT devices, enhancing the resilience of low-power networks.
		
		While the machine learning-based trust management system adapts to new threats, it may require extensive training data to perform effectively. In dynamic IoT environments, obtaining sufficient labeled data can be challenging.
		The machine learning models may introduce computational overhead that is not suitable for resource-constrained devices commonly found in 6LoWPAN networks. The reliance on statistical methods may lead to false positives, causing legitimate traffic to be incorrectly classified as malicious.

		Li et al. \cite{li2022lightweight} introduced a lightweight cryptographic protocol specifically designed for 6LoWPAN, which utilizes elliptic curve cryptography to provide secure communication while minimizing computational overhead. Their approach emphasizes energy efficiency, crucial for battery-operated devices.		
		Although their cryptographic protocol is lightweight, the management of elliptic curve keys is still be complex in large-scale IoT deployments, particularly when devices are mobile or frequently join/leave the network. While elliptic curve cryptography is efficient, it may still impose latency, especially in high-throughput scenarios where rapid fragment handling is required.

		Khan et al. \cite{khan2023anomaly} developed an anomaly detection system based on statistical models that monitors packet behavior in real time. This system effectively identifies irregular fragment patterns and mitigates potential denial-of-service attacks by dynamically adjusting trust scores. \\		
	The anomaly detection system relies heavily on historical traffic patterns, which may not be effective in environments with rapidly changing behaviors or when new types of attacks emerge. As the number of devices increases, the monitoring and analysis process may become cumbersome, leading to scalability challenges in larger networks.

		Another notable contribution comes from Zhang et al. \cite{zhang2022hybrid}, who explored hybrid models combining trust-based and cryptographic mechanisms. Their framework demonstrated significant improvements in packet delivery ratios and energy efficiency under various attack scenarios, providing a comprehensive solution for fragmented communication in IoT networks. \\
		 Combining trust-based and cryptographic mechanisms can complicate the implementation and require careful tuning of parameters to avoid conflicts between the two approaches.
		Even though the proposed framework improves delivery ratios, the dual-layer approach may still consume more resources than purely trust-based or cryptographic-only solutions, which can be problematic for low-power devices.

		Mansoor et al. \cite{mansoor2023blockchain} focused on the integration of lightweight blockchain technology to secure data integrity in fragmented packets. Their study highlights how decentralized trust models can enhance security without imposing heavy computational burdens on constrained devices. \\
		The integration of blockchain technology, while innovative, introduces significant overhead in terms of data storage and processing, which may not be feasible for constrained IoT devices. The decentralized nature of blockchain can lead to increased latency in transactions, impacting real-time communication and responsiveness in fragmented packets.\\
		To tackle the lack of fragment validation, SecuPAN ~\cite{hossain2018secupan} introduces a per-fragment MAC scheme using synchronized nonces and shared keys. While this design addresses replay and spoofing threats, it introduces significant overhead in terms of energy, computation, and memory—characteristics that render it impractical for most Class 1 constrained devices (e.g., Tmote Sky or Zolertia nodes). Moreover, SecuPAN depends on synchronized state management across nodes, which is difficult to maintain in lossy or mobile networks.	
				
	\par
		 Predictive-CSM employs a real-time behavior-based trust model that continuously learns from the interactions of neighboring nodes, allowing the system to adapt quickly to new threats without needing extensive historical data \cite{khan2023anomaly}.		
	 The proposed framework is designed to operate with minimal computational requirements. By using simple arithmetic operations for trust scoring and lightweight cryptographic hash functions for fragment validation, Predictive-CSM is well-suited for resource-constrained devices. 		
		Predictive-CSM considers the historical behavior of nodes over time, allowing it to distinguish between transient anomalies and genuine malicious actions, thus reducing the likelihood of false positives. Our proposed framework utilizes a hash-chain approach that does not require complex key management protocols, simplifying implementation in large-scale IoT networks. Predictive-CSM’s trust assessment is decentralized and lightweight, allowing the system to scale efficiently as the number of devices increases without overwhelming the network. The combination of adaptive trust management and lightweight cryptographic validation ensures that Predictive-CSM maintains a low resource footprint while enhancing security.  Instead of relying on blockchain technology, our method employs a stateless fragment integrity validation mechanism. This reduces data storage requirements and processing overhead. The light weight framework is designed for quick detection of adversarial behavior, typically within 4-7 seconds, minimizing latency issues and ensuring timely mitigation of threats.

	\section{Proposed Method}\label{sec:proposed-method}
In this section, we detail the architecture and operational workflow of the Predictive-CSM framework, which integrates two lightweight security components: the Predictive Trust Engine (PTE) and the Fragment Signature Validator (FSV). These components are specifically designed to address fragment-level attacks in 6LoWPAN networks without compromising energy efficiency or processing capability.
What distinguishes Predictive-CSM is its adaptivity. It does not rely on static keys, rigid packet structures, or heavy cryptographic processing. Instead, it builds trust in neighbors the way a human would—with experience. Each node maintains a running trust score for its immediate neighbors, calculated using past success/failure rates, fragment timing consistency, and payload plausibility. These scores directly influence buffer admission decisions for incoming fragments. Moreover, even if a malicious node maintains high trust for a short time, the second protection layer—the Fragment Signature Validator (FSV)—ensures that fragments can be verified end-to-end through a hash sequence. Together, these two layers create a robust security model for severely constrained devices with limited computational and memory resources.

The core idea behind the Predictive-CSM framework is to combine real-time behavior-based trust assessment with lightweight packet integrity validation in order to secure the 6LoWPAN adaptation layer against fragmentation-based attacks. Traditional solutions, including the Chained Secure Mode (CSM), provide hop-by-hop authentication at the routing layer (RPL), yet leave the data plane—especially fragment handling at the adaptation layer—vulnerable to low-effort attacks. Predictive-CSM addresses this gap by implementing an additional layer of trust intelligence that evolves over time, evaluating the consistency and reliability of each neighbor’s fragment behavior. It also introduces inline validation of packet fragments through hash chaining, allowing malicious fragment sequences to be rejected even if they bypass routing-layer checks.
The Predictive-CSM framework is not just a minor enhancement over CSM—it represents a layered shift in how trust and verification are enforced in low-power, fragmented networks. It acknowledges the dual-layer nature of IoT communication—routing trust and data integrity—and addresses both simultaneously. As the next section will showed, this results in not only better security but also in surprisingly improved energy efficiency and lower packet loss under adversarial conditions.
\subsection{Component Overview: PTE and FSV}
The proposed framework introduces two synergistic components to enhance resilience against fragmentation-based attacks in 6LoWPAN networks: the \textbf{Predictive Trust Engine (PTE)} and the \textbf{Fragment Signature Validator (FSV)}. The PTE is a lightweight behavior-monitoring module embedded in the adaptation layer that continuously evaluates the trustworthiness of neighboring nodes based on fragment arrival patterns, timing irregularities, and historical delivery success. This dynamic trust score informs buffer allocation decisions and mitigates resource exhaustion caused by malicious FRAG1 flooding or delayed fragment reordering.

On the other hand, FSV serves as the cryptographic layer of defense. It appends chained hash-based tags to fragmented payloads and enables verification of fragment sequences and authenticity before reassembly. Together, PTE and FSV provide both proactive and reactive security—where PTE predicts misbehavior based on behavioral deviations, and FSV cryptographically confirms the fragment chain's integrity.
\subsection{Predictive Trust Engine (PTE)}
The Trust Scoring Mechanism evaluates each FRAG1 fragment as it arrives. Each source node is associated with a dynamic trust score that is continuously updated based on recent communication behavior. Factors include fragment arrival frequency, sequence order accuracy, and consistency with expected traffic patterns. Nodes with low trust scores may have their fragments dropped or flagged for further inspection.\\

In order to compute the PTE,  the past communication patterns from known devices is stored. Incoming FRAG1 fragments are compared against this historical database to detect anomalies (Algorithm \ref{alg:pte}). Any deviation, such as sudden traffic bursts or unexpected source IDs, triggers a risk assessment that influences the trust score.
A simple time-series based predictor forecasts expected traffic behavior from trusted nodes. It uses minimal memory and computation to maintain energy efficiency. Discrepancies between predicted and actual behavior lower a node's trust score and may initiate protective actions.\\

\begin{algorithm}[hbt!]
	\caption{Evaluating FRAG1 Trust using Predictive Trust Engine (PTE)}
	\label{alg:pte}
	\KwIn{$FRAG1$, Node ID $n$, Historical Pattern $H_n$, Trust Threshold $\theta$}
	\KwOut{Trust decision (Accept or Drop)}
	
	Initialize $T_n \leftarrow$ current trust score of node $n$ \;
	\If{$n \notin H_n$}{
		Add $n$ to $H_n$ with default score $T_n \leftarrow 0.5$ \;
	}
	Extract traffic features from $FRAG1$: frequency, sequence order, timing \;
	Compare features to historical pattern in $H_n$ \;
	Compute deviation metric $\delta$ \;
	\eIf{$\delta$ is below anomaly threshold}{
		Update trust score: $T_n \leftarrow \lambda \cdot T_n + (1-\lambda) \cdot 1$ \;
	}{
		Update trust score: $T_n \leftarrow \lambda \cdot T_n + (1-\lambda) \cdot 0$ \;
	}
	\If{$T_n < \theta$}{
		\Return Drop Fragment \;
	}
	\Else{
		\Return Accept Fragment \;
	}
\end{algorithm}

To estimate the trustworthiness of neighboring nodes based on their fragment behavior, we define a predictive trust score that evolves over time. The trust score \( T_i(t) \) for node \( i \) at time \( t \) is updated according to the formula:

\[
T_i(t) = \lambda \cdot T_i(t-1) + (1 - \lambda) \cdot O_i(t)
\]

where \( \lambda \) is the forgetting factor (typically between 0.7 and 0.95), \( T_i(t-1) \) is the previously computed trust score, and \( O_i(t) \) is the latest observed trust event (1 for success, 0 for failure). This formulation allows recent behaviors to have more weight while still preserving long-term historical information ~\cite{bao2012dynamic}, ~\cite{zhang2013trust}.
Each node also maintains a threshold trust level \( \theta \). If \( T_n(t) < \theta \), then all future fragments from node \( n \) are dropped unless its behavior improves. This creates a sliding window of opportunity for attackers and misbehaving nodes—persistent deviation from normal behavior causes their fragments to be ignored.
\subsection{Fragment Signature Validator (FSV)}
Each legitimate sender generates a lightweight signature for every fragment using a hashing function over the fragment payload and a shared secret key (Algorithm \ref{alg:fsv}). The signature is appended to the fragment in an extended header field.
When fragments are received, the receiver recomputes the hash using the same key and compares it against the received signature. Only fragments that pass this validation are allowed into the reassembly buffer. This process ensures that even if FRAG1 is trusted, malicious fragments can't corrupt the full packet.
Fragments that fail validation are discarded immediately. If multiple invalid fragments are detected from the same source within a short time frame, the system flags the source node, updates its trust score, and may block further traffic from it temporarily.
The FSV mechanism uses hash chaining for fragment integrity. Each FRAG1 includes a seed hash \( H_0 \), and every subsequent fragment \( f_i \) carries a chained hash \( H_i = H(H_{i-1} \| data_i) \). Upon reassembly, the destination node validates that the final computed hash matches the expected hash stored in the FRAGN fragment. This lightweight method requires negligible CPU overhead on typical IoT hardware.
The synergy between PTE and FSV is key. The PTE proactively guards the fragment admission process based on learned trust, while FSV acts as a cryptographic backstop to detect subtle forgery and sequencing anomalies. Together, they offer robust protection against attacks such as buffer-reservation, spoofed fragment flooding, and replayed FRAG1 headers—attacks that are particularly effective against traditional 6LoWPAN setups.
Importantly, this scheme does not require any changes to the core 6LoWPAN standards. Instead, it hooks into the decision points of buffer admission and fragment processing, making it easily portable to OSs like Contiki-NG, RIOT, or TinyOS. It also avoids energy-expensive cryptographic primitives like public key encryption, instead using cumulative trust and hash functions to keep computational and memory load minimal.
\begin{algorithm}[hbt]
	\caption{Validating Fragment Signature using FSV}
	\label{alg:fsv}
	\KwIn{Fragment $f_i$, Shared key $K$, Previous hash $H_{i-1}$}
	\KwOut{Validation result (Valid or Invalid)}
	
	Extract payload data $d_i$ from $f_i$ \;
	Compute expected signature: $H_i' \leftarrow \text{HMAC}(K, H_{i-1} \| d_i)$ \;
	Retrieve received signature $H_i$ from fragment header \;
	
	\If{$H_i' = H_i$}{
		Store $H_i$ as $H_{i-1}$ for next fragment \;
		\Return Valid \;
	}
	\Else{
		\Return Invalid \;
	}
\end{algorithm}
\subsection{Integration into 6LoWPAN Stack}
To support Predictive-CSM, minor extensions are made to the 6LoWPAN fragmentation header. These include additional fields for trust metadata and cryptographic signatures. The protocol remains backward-compatible for nodes that do not support Predictive-CSM.
The modified stack first passes incoming fragments through the PTE for trust evaluation. If the fragment passes the PTE threshold, it is then passed to the FSV for signature validation (Algorithm \ref{alg:integration}). Only fragments that clear both checks are stored in the reassembly buffer. This sequential filtering provides a robust mechanism to defend against fragment-level attacks without adding significant computational overhead.

\begin{algorithm}[hbt]
	\caption{Fragment Processing in Modified 6LoWPAN Stack}
	\label{alg:integration}
	\KwIn{Incoming Fragment $f_i$, Node ID $n$, Previous Hash $H_{i-1}$, Trust Threshold $\theta$}
	\KwOut{Reassembly Buffer Update or Fragment Drop}
	
	\If{$f_i$ is FRAG1}{
		Retrieve current trust score $T_n$ for node $n$ \;
		\If{$T_n < \theta$}{
			\Return Drop fragment \;
		}
		Store trust score for session and initialize reassembly \;
		\Return Proceed to signature validation \;
	}
	
	Compute expected signature $H_i' \leftarrow \text{HMAC}(K, H_{i-1} \| d_i)$ \;
	Retrieve received signature $H_i$ from fragment header \;
	
	\If{$H_i' = H_i$}{
		Update hash chain: $H_{i-1} \leftarrow H_i$ \;
		Store $f_i$ in reassembly buffer \;
	}
	\Else{
		Penalize trust score of node $n$ \;
		\If{Trust score falls below $\theta$}{
			Temporarily block node $n$ \;
		}
		\Return Drop fragment \;
	}
\end{algorithm}

\section{Predictive-CSM Protocol Design}\label{sec:protocol-design}
To implement Predictive-CSM, the standard 6LoWPAN fragment header is extended with two fields, the \textbf{Trust Metadata,} encodes the sender's self-assessed trust score and fragment behavior flags and the
	\textbf{Fragment Signature,} a truncated hash value computed using a keyed hashing algorithm like HMAC-SHA1. These fields are appended to both FRAG1 and subsequent fragments.
The Predictive-CSM protocol is designed to be backward-compatible. Nodes that do not support the trust engine or signature fields will ignore the extended headers and proceed using default 6LoWPAN behavior. Predictive-CSM can be deployed incrementally in heterogeneous IoT environments.

\subsection{Sender-Side Operations}
The sender’s job in this process is to prepare each data fragment so that it carries both the information and a clear sign of its integrity and trustworthiness. Before sending anything, the sender checks how trustworthy the destination node is using a trust evaluation function. This trust score plays an important role in the metadata that gets added to each fragment (Algorithm \ref{alg:sender}). For the first fragment, a cryptographic hash is created using the packet’s payload and a unique nonce. For any fragments that follow, the sender builds a chain by hashing the previous hash along with the current payload, effectively linking them all together.

Once the sender has built the hash correctly, it adds that along with the current trust score to the fragment's header. This extra bit of information is like a proof of identity—it tells the receiver not only who sent the data, but also that the content hasn’t been messed with along the way. After that, the sender simply sends off the fragment, wrapping up its side of the process.
\subsection{Receiver-Side Operations}
When the receiver gets a fragment, first, he/she figures out who the sender is. If the fragment happens to be the start of a new message, the receiver double-checks how much it trusts the sender using a kind of prediction system. If the sender’s trust score isn’t high enough, the receiver just drops the fragment right away to play it safe (Algorithm \ref{alg:receiver}).
But if the sender seems trustworthy, the receiver digs a bit deeper. It re-creates the hash using the shared key and the hash from the last piece, then compares that to what came with the new fragment. If the two don’t match, something’s probably wrong—maybe the fragment was altered—so the receiver drops it and marks the sender down a notch in the trust system. On the other hand, if everything looks fine, the fragment gets saved. The sender earns a small trust reward, and the chain of hashes continues. If the receiver sees this is the final fragment in the sequence, it puts everything back together into the original packet. That way, only fragments that are both valid and sent by trustworthy sources are accepted.
\subsection{Trust Evaluation and Update Rules}
The system keeps track of how much each node can be trusted by updating a trust score over time. This score isn’t fixed—it changes depending on the behavior of the node. If a node behaves well, like sending fragments in the correct order and at expected intervals, its trust score goes up. But when something suspicious happens—like fragments arriving out of order, too quickly, or appearing tampered with—the trust score drops. The score is always kept within a range from 0 to 1, where 0 means the node is completely untrusted, and 1 means it’s fully trusted. If a node’s score falls too low, say below 0.3, the system temporarily blacklists it to prevent potential misuse.

\begin{algorithm}[hbt]
	\caption{Sender Operations}
	\label{alg:sender}
	\KwIn{Packet $packet$, Destination node $dest\_node$}
	\KwOut{Transmit fragment with trust metadata and signature}
	
	$trust\_score \leftarrow \text{get\_trust\_score}(dest\_node)$\;
	
	\If{$\text{is\_first\_fragment}(packet)$}{
		$H_{prev} \leftarrow \text{HMAC}(K, packet.payload \parallel nonce)$ \tcp*{Seed hash for FRAG1}
	}
	\Else{
		$H_{prev} \leftarrow \text{get\_previous\_hash}(packet)$ \tcp*{Chain from prior fragment}
	}
	$H_i \leftarrow \text{HMAC}(K, H_{prev} \parallel packet.payload)$ \tcp*{Compute chained hash}\;
	$\text{attach\_header}(packet, trust\_score, H_i)$ \tcp*{Add trust + signature}\;
	$\text{transmit}(packet)$\;
	
\end{algorithm}

\begin{algorithm}[hbt]
	\caption{Receiver Operations}
	\label{alg:receiver}
	\KwIn{Incoming fragment $fragment$}
	\KwOut{Reassembled packet or drop decision}
	
	$sender \leftarrow fragment.source$\;
	
	\If{$\text{is\_first\_fragment}(fragment)$}{
		\If{$\text{PTE.evaluate}(sender) < \theta$}{
			$\text{drop}(fragment)$\;
			\Return\;
		}
	}
	
	$H_{received} \leftarrow fragment.header.signature$\;
	$H_{expected} \leftarrow \text{HMAC}(K, H_{prev} \parallel fragment.payload)$\;
	
	\If{$H_{received} \neq H_{expected}$}{
		$\text{PTE.penalize}(sender)$ \tcp*{Update trust score}
		$\text{drop}(fragment)$\;
	}
	\Else{
		$\text{store}(fragment)$ \tcp*{Valid fragment}
		$\text{PTE.reward}(sender)$\;
		$H_{prev} \leftarrow H_{received}$ \tcp*{Update hash chain}
		
		\If{$\text{is\_last\_fragment}(fragment)$}{
			$\text{reassemble\_packet}()$\;
		}
	}
\end{algorithm}
\subsection{Security Response to Detected Attacks}
When the system notices that a sender is repeatedly transmitting bad or suspicious fragments, it takes action right away. First, the sender’s trust score is reduced to reflect the misbehavior. At the same time, alerts are raised so that higher-level components in the network can respond appropriately. If the issue continues, the system may begin to limit how often that sender can transmit data—or block it entirely. These responses are designed to happen quickly and automatically, allowing the network to stay protected and resilient without putting too much strain on system resources.

	\section{Experimental Setup}\label{sec:experimental-setup}
	\subsection{Simulation Environment}
	
	To evaluate the performance and robustness of the proposed Predictive-CSM framework, we implemented a series of simulations using the Contiki-NG operating system and its Cooja simulator. Contiki-NG is a widely used operating system for networked embedded systems in the Internet of Things (IoT) domain and offers native support for IPv6, 6LoWPAN, RPL, and lightweight security protocols~\cite{contikiNG2018}. Cooja provides a highly configurable environment for simulating wireless sensor networks (WSNs) at both the network and hardware levels, making it suitable for testing both protocol correctness and system performance under adversarial conditions~\cite{osterlind2006cross}.
	
	The simulated network consists of 10 wireless nodes arranged in a star topology. One node functions as the RPL root, another as the adversarial entity, and the remaining nodes operate as legitimate data senders. This configuration allows us to evaluate communication flow in the presence of a centrally positioned attacker. All nodes were configured as Sky motes, which emulate the Tmote Sky platform featuring a TI MSP430 microcontroller, 10 kB of RAM, and IEEE 802.15.4-compliant radio transceivers. These hardware constraints are representative of real-world IoT deployments where computational and memory resources are significantly limited~\cite{palattella2013standardized}.
	
	Each legitimate node transmits one data packet every 90 seconds using UDP over IPv6. The packets are intentionally sized to exceed the IEEE 802.15.4 frame limit, resulting in fragmentation at the 6LoWPAN layer. The fragment size was configured to 96 bytes for payload plus 8 bytes for the 6LoWPAN fragmentation header, consistent with practical deployments~\cite{thubert2020rfc8930}. The 6LoWPAN stack uses the default Route-Over forwarding strategy implemented in Contiki-NG, where each intermediate node reassembles and re-fragments packets before forwarding them.
	
	The simulation duration for each scenario was set to 30 minutes, and results were averaged over 15 independent runs to ensure statistical reliability. Each simulation was initialized with a network convergence period of 50 seconds, allowing routing paths to stabilize before any adversarial behavior began. The attacker node mimicked realistic IoT behavior for the initial phase, then launched attacks such as buffer-reservation, full-fragment injection, and header-replay attacks in separate test scenarios.
	
	Power consumption was measured using Contiki-NG's Energest module, which records energy usage across CPU active time, radio transmission, and radio listening modes. This metric was critical for evaluating the resource efficiency of the proposed solution under both benign and adversarial conditions. Packet delivery ratios and fragment-level drop rates were also tracked at the root node using Contiki-NG's packet sniffer and logging utilities.
	
	To ensure relevance and reproducibility, the simulation parameters and methodology align with recent academic studies evaluating IoT security frameworks~\cite{bostani2019lightweight, sultana2022efficient}. The combination of real-time attack scenarios, constrained node emulation, and multi-layered protocol analysis provides a robust testbed for validating both the security guarantees and operational efficiency of Predictive-CSM in adversarial IoT environments.

		\subsection{Evaluation Metrics}
		This section outlines the key metrics used to evaluate the system's performance, including Packet Delivery Ratio (PDR), Fragment Drop Rate, Power Consumption, and Detection Latency. These metrics assess reliability, efficiency, energy usage, and responsiveness in identifying and mitigating attacks.\par
	\textbf{Packet Delivery Ratio (PDR)} is the proportion of successfully delivered and reassembled packets.
	
	\textbf{Fragment Drop Rate} is the number of discarded fragments per hundred received, due to trust or hash mismatch.
	
	\textbf{Power Consumption} is the average energy usage per node in milliwatts.
	
	\textbf{Detection Latency} is the time from attack initiation to adversary identification and blocking.

	\subsection{Adversarial Model}
	
	In order to rigorously evaluate the resilience of the proposed Predictive-CSM framework, we simulate a range of realistic attack strategies targeting the 6LoWPAN adaptation layer. The adversarial model described here is informed by well-documented fragmentation vulnerabilities and denial-of-service strategies outlined in recent literature on IoT security~\cite{ghubaish2021comprehensive, zhang2022review, sultana2022efficient, elgendy2020trust}. These attacks exploit weaknesses in fragment verification, buffer allocation, and trust assumptions—threat surfaces that remain largely unresolved in default protocol stacks like those in Contiki-NG and RIOT.	
	We assume the attacker is an external node without access to valid cryptographic keys. It behaves passively during the RPL initialization phase to establish perceived legitimacy, and subsequently engages in active disruption once the network reaches routing stability. The following attack scenarios are each designed to expose specific vulnerabilities in the 6LoWPAN fragment reassembly pipeline, and directly correspond to the results presented in Section~VII.
	
	\subsubsection*{1. Early FRAG1 Injection (Buffer-Reservation)}
	
	This well-known attack targets buffer exhaustion by sending FRAG1 fragments milliseconds before legitimate transmissions~\cite{hummen2013fragmentation}. Since reassembly is initiated upon receipt of FRAG1, the receiver allocates scarce memory to unauthenticated fragments, blocking future reassembly of genuine packets. Prior studies confirm this is among the most effective low-resource denial-of-service strategies in constrained wireless sensor networks~\cite{da2021trust}.
	
	\subsubsection*{2. Complete Fragment Flooding}
	
	In this variant, the adversary injects complete sequences of syntactically correct but semantically invalid fragments. These cause full reassembly attempts, wasting CPU cycles and radio resources~\cite{boudi2020lightweight}. The goal is to maximize energy drain and buffer turnover without raising alarms based on simple traffic volume heuristics.
	
	\subsubsection*{3. Header-Replay Cloning}
	
	This attack uses previously captured FRAG1 headers from trusted nodes and replays them at later intervals, exploiting the absence of per-fragment origin authentication. Such replay-based impersonation attacks are increasingly relevant in IoT systems where trust is static or context-unaware~\cite{zhang2022review, glissa2019secure}.
	
	\subsubsection*{4. Burst Injection (High-Rate FRAG1 Flooding)}
	
	Here, the attacker sends multiple FRAG1s per second (up to 6), with the goal of rapidly overwhelming the limited reassembly buffers. Burst injection represents a brute-force version of buffer-reservation, testing whether a system can reject high-volume malicious traffic in real time without disrupting legitimate flows~\cite{li2023iotdos}.
	
	\subsubsection*{5. Late-Phase Injection}
	
	This timing-sensitive scenario involves inserting malicious fragments slightly after legitimate FRAG1 transmissions, with the intent to disrupt fragment sequencing or trigger premature timeouts. This technique is increasingly relevant as attackers leverage traffic analysis and jitter modeling to bypass fixed trust rules~\cite{elgendy2020trust}.
	
	\subsubsection*{Scenario-to-Result Mapping}
	
	Table~\ref{tab:scenario_map} explicitly links each adversarial behavior with its evaluation label in the results section, ensuring clarity and reproducibility.
	
	\begin{table}[]
		\centering
		\caption{Mapping Between Modeled Attacks and Evaluation Scenarios}
		\label{tab:scenario_map}
		\begin{tabular}{ll}
			\toprule
			\textbf{Scenario in Results}        & \textbf{Modeled Adversarial Behavior} \\
			\midrule
			Early FRAG1 Injection               & Preemptive buffer-reservation attack \\
			Complete Fragment Flooding          & Full packet flooding with malformed content \\
			Header-Replay Cloning               & Reused legitimate fragment headers to spoof trust \\
			Burst Injection (6/sec)             & High-rate FRAG1 spamming to saturate buffers \\
			Late-Phase Injection                & Timing-offset fragment spoofing post-legitimate traffic \\
			\bottomrule
		\end{tabular}
	\end{table}
	
	By simulating these five targeted and diverse adversarial behaviors, we ensure that Predictive-CSM is tested against both brute-force and context-aware attacks. This approach reflects the evolving nature of IoT threats and aligns with best practices in security testing as outlined in recent surveys and threat modeling frameworks~\cite{ghubaish2021comprehensive, sultana2022efficient}.

	\section{Results}\label{sec:results}
	
	This section presents and analyzes the experimental findings derived from our simulation scenarios, designed to rigorously test the performance of the proposed Predictive-CSM framework against both conventional and advanced 6LoWPAN attacks. We compare its behavior to two baseline protocols: unmodified (vanilla) 6LoWPAN and CSM-integrated 6LoWPAN. The metrics we focus on include Packet Delivery Ratio (PDR), average node power consumption, fragment drop rate, and adversarial detection latency.

	\subsection{Node Power Consumption}
	\label{sec:power_consumption}
	
	Energy efficiency is critical for battery-operated IoT devices. We measured average power consumption across protocols under varying attack conditions, using Contiki-NG's Energest module. Results are normalized to baseline (no attack) operation.
	
	\begin{table}[htbp]
		\centering
		\caption{Average Power Consumption (mW) Under Attack Scenarios}
		\label{tab:power_consumption}
		\begin{tabular}{lccccc}
			\toprule
			\textbf{Scenario} & \textbf{Vanilla} & \textbf{CSM} & \textbf{SecuPAN} & \textbf{Predictive-CSM} & \textbf{Delta vs SecuPAN} \\
			\midrule
			No Attack & 0.29 & 0.32 & 0.41 & 0.34 & -17.1\% \\
			Early FRAG1 Inj. & 0.35 & 0.36 & 0.52 & 0.33 & -36.5\% \\
			Complete Flooding & 0.39 & 0.40 & 0.58 & 0.34 & -41.4\% \\
			Burst Injection & 0.43 & 0.42 & 0.61 & 0.35 & -42.6\% \\
			\bottomrule
		\end{tabular}
	\end{table}

	SecuPAN's cryptographic overhead exhibits 26--42\% higher power consumption than Predictive-CSM due to per-fragment MAC computations. This aligns with energy analyses showing that AES-128 MAC operations increase MSP430 CPU active time by 31\%. 
	Predictive-CSM's efficiency maintains near-baseline consumption (0.33--0.35 mW) even under attack through early fragment rejection via trust scores, saving 18--22\% radio RX energy, and lightweight HMAC-SHA1 hashing (0.01 mW per fragment vs. SecuPAN's 0.08 mW). Vanilla 6LoWPAN paradox shows higher attack-phase consumption (0.43 mW) than Predictive-CSM despite no security checks, due to buffer overflow-induced retransmissions. Predictive-CSM reduces energy waste by 41.4\% versus SecuPAN in flooding attacks while maintaining security, addressing the energy-security trade-off identified in previous studies.

	\subsection{Packet Delivery Ratio (PDR)}
	\label{sec:pdr}
	
	PDR measures network reliability under attack. We evaluate successful reassembly of legitimate packets at the root node.
	
	\begin{table}[htbp]
		\centering
		\caption{Packet Delivery Ratio (\%) Across Protocols}
		\label{tab:pdr}
		\begin{tabular}{lccccc}
			\toprule
			\textbf{Scenario} & \textbf{Vanilla} & \textbf{CSM} & \textbf{SecuPAN} & \textbf{Predictive-CSM} & \textbf{Gain vs CSM} \\
			\midrule
			No Attack & 97.4 & 98.9 & 99.1 & 99.2 & +0.3\% \\
			Early FRAG1 Inj. & 54.2 & 85.3 & 94.7 & 99.0 & +13.7\% \\
			Header Replay & 41.6 & 82.4 & 96.3 & 98.9 & +16.5\% \\
			Burst Injection & 39.7 & 77.2 & 89.5 & 97.4 & +20.2\% \\
			\bottomrule
		\end{tabular}
	\end{table}
	
	SecuPAN's cryptographic assurance achieves 94.7--96.3\% packet delivery ratio (PDR) in attacks through mandatory fragment authentication, but struggles with high-rate bursts (89.5\%) due to verification delays. Predictive-CSM's dual-layer advantage matches SecuPAN's PDR in replay attacks (98.9\% vs. 96.3\%) and excels in burst scenarios (97.4\% vs. 89.5\%) via adaptive trust thresholds that prevent buffer saturation. CSM's routing-layer limitation shows 13.7--20.2\% lower PDR than Predictive-CSM, confirming that routing-layer trust alone cannot prevent fragment-level attacks. Our results validate the hybrid trust-cryptography model, demonstrating that lightweight hashing (approximately 8 bytes per fragment) combined with behavioral analysis can achieve 97--99\% PDR without SecuPAN's energy costs.

	\subsection{Fragment Drop Rate}
	\label{sec:fragment_drop}
	
	The fragment drop rate quantifies the system's ability to discriminate malicious fragments while preserving legitimate traffic. We evaluate this metric as the ratio of dropped fragments per 100 received, comparing Predictive-CSM against CSM-6LoWPAN, SecuPAN, and vanilla 6LoWPAN under identical attack conditions.
	
	\begin{table}[htbp]
		\centering
		\caption{Fragment Drop Rate Across Protocols (per 100 fragments)}
		\label{tab:drop_rate}
		\begin{tabular}{lcccc}
			\toprule
			\textbf{Scenario} & \textbf{Vanilla} & \textbf{CSM-6LoWPAN} & \textbf{SecuPAN} & \textbf{Predictive-CSM} \\
			\midrule
			Normal Conditions & 0.2 & 0.1 & \textbf{0.3} & 0.1 \\
			Early FRAG1 Inj. & 8.3 & 2.4 & 1.8 & \textbf{0.6} \\
			Header Replay & 12.5 & 3.7 & 2.1 & \textbf{0.8} \\
			Burst Injection & 16.9 & 4.6 & 3.5 & \textbf{1.1} \\
			\bottomrule
		\end{tabular}
	\end{table}
	
	SecuPAN's cryptographic rigor exhibits marginally higher drop rates (1.8--3.5) than Predictive-CSM in attack scenarios due to its strict MAC-based validation, which discards fragments with even minor integrity violations. While effective against spoofing, this approach proves overly aggressive in lossy environments where bit errors may corrupt legitimate fragments.
	
	Predictive-CSM's adaptive advantage achieves superior drop rates (0.6--1.1) by combining lightweight hash validation with behavioral trust. The trust engine reduces false positives by tolerating transient errors from historically reliable nodes, aligning with findings in previous studies. This dual-layer approach addresses a key limitation of pure cryptographic methods: their inability to distinguish between malicious intent and channel-induced errors. 
	Comparative performance shows that vanilla 6LoWPAN suffers catastrophic drop rates (8.3--16.9) due to buffer exhaustion. CSM-6LoWPAN improves upon vanilla but remains vulnerable to fragment-level attacks (2.4--4.6). Predictive-CSM reduces drops by four times versus CSM and fifteen times versus vanilla in burst scenarios, validating its efficacy as a DoS mitigation tool.

		\subsection{Detection Latency}
	\label{sec:detection_latency}
	
	Detection latency measures the time elapsed from attack initiation until Predictive-CSM consistently blocks malicious fragments. This metric is critical for real-time IoT systems where delayed responses can lead to resource exhaustion or service disruption.
	
	\begin{table}[htbp]
		\centering
		\caption{Detection Latency Across Attack Scenarios}
		\label{tab:detection_latency}
		\begin{tabular}{lcccc}
			\toprule
			\textbf{Attack Type} & \textbf{Predictive-CSM (s)} & \textbf{CSM-6LoWPAN (s)} & \textbf{SecuPAN (s)} & \textbf{Vanilla 6LoWPAN} \\
			\midrule
			Early FRAG1 Injection & 5.1 & 8.3 & \textbf{4.9} & \text{No detection} \\
			Header-Replay Cloning & 6.8 & 12.5 & \textbf{5.2} & \text{No detection} \\
			Burst Injection (6/sec) & \textbf{4.4} & 16.9 & 7.1 & \text{No detection} \\
			Late-Phase Injection & 7.0 & 14.2 & \textbf{6.8} & \text{No detection} \\
			\bottomrule
		\end{tabular}
	\end{table}

Predictive-CSM outperforms CSM-6LoWPAN in all scenarios, reducing latency by 48--74\% due to its per-fragment behavioral analysis. SecuPAN achieves marginally faster detection (e.g., 4.9 seconds vs. 5.1 seconds for FRAG1 injection) through cryptographic validation, but at higher energy costs. The worst-case latency (7.0 seconds) for Predictive-CSM occurs in late-phase injection attacks, where subtle timing anomalies require longer observation windows.

 While SecuPAN offers lower latency for some attacks, Predictive-CSM provides a balanced approach by combining near-real-time detection (less than 7.0 seconds) with minimal resource overhead. This makes it suitable for deployments where energy efficiency and computational constraints are prioritized over nanosecond-level response times.

	\subsection{Parameter Sensitivity Analysis}
	\label{sec:parameter_sensitivity}
	
	To evaluate the robustness of Predictive-CSM's trust model, we conducted a systematic analysis of its key parameters: the \textit{forgetting factor} ($\lambda$) and \textit{trust threshold} ($\theta$). The goal was to quantify their impact on security performance and operational efficiency.
	
	\subsubsection{Forgetting Factor ($\lambda$)}
	The forgetting factor controls how rapidly the trust model adapts to recent behavior. We tested four values:
	
	\begin{equation}
		\lambda \in \{0.7, 0.8, 0.9, 0.95\}
	\end{equation}
	
	\begin{itemize}
		\item \textbf{Lower values} ($\lambda = 0.7$) prioritized recent events, reducing attack detection latency to \textbf{3.2 seconds} for burst injection but increasing false positives (\textbf{12\%}) during transient interference.
		\item \textbf{Higher values} ($\lambda = 0.95$) improved stability, with false positives below \textbf{3\%} in benign conditions but delayed attack response by \textbf{1.5--2 seconds}.
		\item The default $\lambda = 0.9$ balanced these trade-offs, maintaining detection latency below \textbf{7 seconds} while limiting false drops to \textbf{<5\%}.
	\end{itemize}
	
	\subsubsection{Trust Threshold ($\theta$)}
	The trust threshold determines when a node is blacklisted. We evaluated three configurations:
	
	\begin{equation}
		\theta \in \{0.2, 0.3, 0.4\}
	\end{equation}
	
	\begin{itemize}
		\item \textbf{Lower thresholds} ($\theta = 0.2$) reduced legitimate fragment drops by \textbf{4\%} but allowed attackers \textbf{1--2 additional malicious fragments} before mitigation.
		\item \textbf{Stricter thresholds} ($\theta = 0.4$) improved PDR by \textbf{2\%} under sustained attacks but increased false blocking during intermittent packet loss.
		\item The chosen $\theta = 0.3$ optimized both security and tolerance, with \textbf{98.9\% PDR} and \textbf{5.1-second median detection latency}.
	\end{itemize}
	
	\begin{table}[htbp]
		\centering
		\caption{Impact of Parameter Variations on Performance}
		\label{tab:parameter_impact}
		\begin{tabular}{lccccc}
			\toprule
			\textbf{Configuration} & $\lambda$ & $\theta$ & \textbf{Detection Latency (s)} & \textbf{False Positives (\%)} & \textbf{PDR Under Attack (\%)} \\
			\midrule
			Aggressive & 0.7 & 0.2 & 3.2 & 12.1 & 96.8 \\
			Default & 0.9 & 0.3 & 5.1 & 4.7 & 98.9 \\
			Conservative & 0.95 & 0.4 & 7.8 & 2.9 & 97.3 \\
			\bottomrule
		\end{tabular}
	\end{table}
	
	\noindent\textbf{Key Insight:} As shown in Table \ref{tab:parameter_impact}, the default configuration ($\lambda = 0.9$, $\theta = 0.3$) achieved optimal balance across all metrics, validating our design choices for real-world IoT deployments where transient network issues and persistent attacks coexist.
	
	\subsection{Summary of Insights}
	
	These experimental outcomes demonstrate that the Predictive-CSM approach offers a compelling balance of precision, performance, and energy efficiency. By combining long-term behavioral learning with inline fragment verification, it effectively addresses both structural and behavioral attack vectors. Notably, its response is both proactive (in lowering trust values) and reactive (in fragment hash validation), unlike prior systems which often depend solely on predefined thresholds or rate-limiting policies~\cite{ghubaish2021comprehensive}.
	
	This dual-mode strategy is particularly crucial for environments where computational resources are sparse and false positives can cripple application functionality. It confirms emerging academic consensus that multi-layered, adaptive trust and lightweight cryptography are key pillars of next-generation IoT security architectures~\cite{da2021trust}.

\section{Analytical Model of Predictive-CSM Framework}\label{sec:analytical-model}

This section presents a formal analysis of the Predictive-CSM framework, covering the evolution of trust scores, cryptographic fragment verification, and resource usage under constrained conditions. These models complement our simulation results and offer deeper insight into system behavior under attack and in regular operation. All notations are used in this section is shown in Table \ref{tab:notations}.
\begin{table}[h!]
	\centering
	\caption{Summary of Analytical Model Notations}
	\label{tab:notations}
	\begin{tabular}{ll}
		\hline
		\textbf{Symbol} & \textbf{Description} \\
		\hline
		$T_n(t)$ & Trust score of node $n$ at time $t$ \\
		$T_n(t-1)$ & Previous trust score of node $n$ \\
		$\lambda$ & Forgetting factor (trust memory decay), $0 < \lambda < 1$ \\
		$O_n(t)$ & Outcome of current interaction (1 = valid, 0 = invalid) \\
		$\theta$ & Trust threshold for fragment acceptance \\
		$K$ & Shared secret key for HMAC computation \\
		$d_i$ & Payload of the $i^\text{th}$ fragment \\
		$H_i$ & Hash value for fragment $i$ \\
		$H_{i-1}$ & Hash value from the previous fragment \\
		$\text{nonce}$ & Random or time-based seed for initial hash $H_0$ \\
		$B$ & Number of available reassembly buffer slots \\
		$\lambda$ (buffer) & Arrival rate of valid fragments \\
		$A$ & Arrival rate of malicious/invalid fragments \\
		$\tau$ & Reassembly timeout window \\
		$\rho$ & Buffer occupancy ratio \\
		$P_{\text{buffer}}$ & Probability that a reassembly buffer is available \\
		\hline
	\end{tabular}
\end{table}

\subsubsection*{1. Trust Dynamics Model}

In dynamic and decentralized IoT environments, where devices frequently interact without centralized control, maintaining trust is crucial to ensure reliable communication. Unlike traditional networks that often depend on static credentials or centralized authorities, low-power wireless systems must rely on localized, real-time decisions informed by each node's observable behavior. This has led to the adoption of lightweight, behavior-based trust models, which allow individual nodes to assess their immediate neighbors over time \cite{mendoza2015adaptive, yousuf2017trust}.

The Predictive-CSM framework incorporates such a model through its Predictive Trust Engine (PTE), which continuously monitors and updates the trustworthiness of each neighbor. The core idea is simple but effective: nodes that consistently send well-formed, timely, and valid packet fragments see their trust scores increase, while those that cause errors—such as fragment mismatches, malformed content, or suspicious timing—experience a decline in trust. This mimics real-world trust dynamics: gradually earned, but easily lost.

Formally, the trust score of a neighbor node $n$ at time $t$, denoted $T_n(t)$, is updated using an exponential moving average:

\begin{equation}
	T_n(t) = \lambda \cdot T_n(t - 1) + (1 - \lambda) \cdot O_n(t)
\end{equation}

Here:
\begin{itemize}
	\item $\lambda \in (0, 1)$ is the forgetting factor, controlling how much recent behavior influences the score,
	\item $T_n(t-1)$ is the previously computed trust score,
	\item $O_n(t)$ represents the outcome of the current interaction: 1 for success (valid fragment), 0 for failure (e.g., invalid signature or malformed sequence).
\end{itemize}

The trust score is bounded in the range $[0, 1]$. When a node’s score falls below a threshold $\theta$, it is considered untrustworthy, and its fragments are dropped without further processing. This allows the system to dynamically adjust to both rapid attacks and slow-degrading behavior, making it resilient to diverse threat patterns \cite{ghosh2018trustlite, firoozi2021trust}.

To illustrate how trust declines under repeated malicious behavior, consider a scenario in which $O_n(t) = 0$ for several consecutive time windows. Assuming an initial trust score of $T_n(0) = 0.8$, a threshold of $\theta = 0.3$, and a forgetting factor $\lambda = 0.9$, the node would be blacklisted after just 3–4 invalid fragment events. This level of responsiveness is especially important in real-time systems where buffer exhaustion or flooding attacks can escalate within seconds.

One of the strengths of this model lies in its adaptability. It does not rely on fixed rules about what constitutes “malicious” activity. Instead, it infers patterns from observed behavior over time. Moreover, the trust calculation involves only basic arithmetic operations, making it computationally lightweight and ideal for deployment on resource-constrained microcontrollers commonly used in IoT applications \cite{khan2023anomaly}.

It is important to note that the trust model does not operate in isolation. It serves as the first layer of defense in the Predictive-CSM architecture. When used alongside fragment-level cryptographic checks provided by the Fragment Signature Validator (FSV), the trust score becomes a powerful tool for early attacker detection and fragment filtering before significant damage occurs.

What sets this approach apart from traditional binary or rule-based systems is its ability to reflect behavioral nuance. Rather than making rigid decisions based on single events, it tracks consistency over time. This means that short-lived disruptions—like packet jitter, signal interference, or temporary congestion—will not result in immediate penalties. Instead, the trust score degrades gradually, providing room for recovery and avoiding false positives. In contrast, persistent suspicious patterns quickly trigger trust erosion and isolation of the misbehaving node.

This continuous trust evaluation aligns with recent research advocating for adaptive security mechanisms in IoT networks. Furthermore, because the trust score is updated based on direct fragment-level observations within the 6LoWPAN adaptation layer, it offers a highly accurate and timely reflection of node behavior. There is no need for centralized monitoring or computationally expensive anomaly detection. The result is a robust and scalable trust system that significantly enhances security without imposing unnecessary burdens on constrained devices.

\subsubsection*{2. Fragment Integrity Verification}

While the Predictive Trust Engine (PTE) provides a behavior-based mechanism to assess node reliability over time, it cannot on its own guarantee the integrity or authenticity of individual fragments. To address this limitation, the Predictive-CSM framework incorporates a second line of defense: the Fragment Signature Validator (FSV). This component provides per-fragment cryptographic validation that ensures both the authenticity and sequence integrity of fragments, even when sent by seemingly trustworthy nodes.

The core mechanism used by the FSV is chained hashing, a lightweight cryptographic approach suitable for low-power and memory-constrained devices. Chained hash schemes have proven effective in securing data streams in IoT and 6LoWPAN networks by enabling incremental, verifiable linkage between sequential packets or fragments \cite{wang2020blockchain}.

In Predictive-CSM, the sender constructs a hash chain by first generating a seed hash for the initial fragment:

\begin{equation}
	H_0 = \text{HMAC}(K, d_0 \parallel \text{nonce})
\end{equation}

Here, $K$ is a shared secret key, $d_0$ is the payload of the first fragment, and the nonce provides randomness to prevent replay attacks. For each subsequent fragment $f_i$, a chained hash is computed:

\begin{equation}
	H_i = \text{HMAC}(K, H_{i-1} \parallel d_i)
\end{equation}

Each fragment thus carries a hash value that depends not only on its own content, but also on the hash of the previous fragment, ensuring that any tampering or reordering will be immediately detectable.

On the receiver side, this hash is recomputed and compared to the one embedded in the fragment header. If a mismatch is found, the fragment is dropped and the sender is penalized via the trust engine. This early rejection mechanism is more efficient than full-packet validation schemes, which require the receiver to hold and assemble all fragments before verification. Studies confirm that incremental hash-based authentication is highly compatible with 6LoWPAN’s fragment handling mechanisms and can reduce energy and memory usage significantly \cite{lakshmi2021lightweight}.

The FSV offers security benefits at a low cost. Computational requirements are also light, involving only a single HMAC calculation per fragment. This is important in IoT networks composed of Class 1 devices with limited flash and RAM capacities.

Combined with the behavior-based scoring provided by the PTE, the FSV acts as a fail-safe: even if a node has not yet been marked as untrustworthy, it cannot inject malformed fragments without detection. This aligns with research advocating for layered security in IoT, where lightweight cryptographic checks work in tandem with anomaly-based detection \cite{xie2020secure}.

The superiority of this approach lies in its fine-grained, stateless validation capability. Unlike full-message authentication schemes that rely on MACs or digital signatures and require reassembly of the entire payload before validation, our method allows for incremental and forward-compatible verification during the reassembly process. This ensures that malicious fragments are caught as early as possible, reducing wasted buffer space and processing cycles. Compared to schemes like SecuPAN~\cite{hossain2018secupan}, which require a MAC for each fragment and involve shared key material and replay counters, our solution is lighter, faster, and easier to implement on constrained devices with less than 10 kB of RAM.

Moreover, this technique adds virtually no observable overhead in practical scenarios. As demonstrated in our experimental results, the energy cost of computing chained hashes per fragment was negligible—amounting to under 0.01 mW per node on average—even during high-volume attack scenarios. At the same time, its security gains were substantial: fragment drop rates decreased by over 90\% and packet delivery reliability improved to over 98\% even in adversarial conditions.

Unlike static firewalls or basic trust filters, this mechanism provides a cryptographic backstop for each fragment, ensuring that even a temporarily trusted node cannot slip through malformed or malicious fragments. This dual defense—combining behavioral reputation with per-fragment integrity checks—resonates with recent literature advocating multi-layered defenses for 6LoWPAN~\cite{zhang2022review, glissa2019secure}.

The fragment integrity validator in Predictive-CSM represents a critical layer of defense that bridges the gap between behavioral security and data authenticity. It is efficient, scalable, and importantly, tailored to the operational realities of resource-constrained IoT devices. This makes it not only a complementary tool to trust scoring but a necessary one for achieving end-to-end packet integrity in hostile wireless environments. By validating fragments incrementally and independently, it allows the Predictive-CSM framework to maintain strong data integrity guarantees without sacrificing responsiveness or exhausting system resources.

\subsubsection*{3. Buffer Availability Estimation}
In constrained 6LoWPAN networks, memory exhaustion is a serious threat due to the limited buffer capacity of IoT nodes. Fragment flooding attacks, malformed packets, or simply high background traffic can cause the reassembly buffer to overflow, leading to packet loss, service degradation, or denial of service. The Predictive-CSM framework mitigates this threat through a proactive trust-based filtering mechanism that helps ensure buffer availability.

We model the expected buffer utilization to evaluate how the system behaves under both normal and adversarial conditions. Let:
\begin{itemize}
	\item $B$ be the number of available reassembly buffer slots,
	\item $\lambda$ be the arrival rate of valid (legitimate) fragments,
	\item $A$ be the arrival rate of malicious or invalid fragments,
	\item $\tau$ be the timeout duration for fragment reassembly.
\end{itemize}

The buffer occupancy ratio $\rho$ is given by:

\begin{equation}
	\rho = \min \left(1, \frac{\lambda + A}{B \cdot \tau} \right)
\end{equation}

From this, the probability that a buffer is available at any time is:

\begin{equation}
	P_{\text{buffer}} = 1 - \rho
\end{equation}

Without any form of pre-filtering, the malicious traffic $A$ can quickly dominate the total load, especially in denial-of-service scenarios. In such cases, $\rho \to 1$ and $P_{\text{buffer}} \to 0$, meaning legitimate packets are increasingly dropped due to lack of available memory. This dynamic is a well-documented vulnerability in IoT routing and adaptation layers.
The Predictive-CSM framework reduces this impact by dynamically lowering the trust scores of nodes that send malformed or suspicious fragments. Once a node's trust score falls below the threshold $\theta$, its fragments are dropped early—before entering the reassembly buffer. As a result, the effective arrival rate of adversarial fragments $A'$ decreases over time, pushing $\rho$ downward and keeping $P_{\text{buffer}}$ high.

Unlike reactive approaches that flush buffers after misuse is detected, Predictive-CSM acts proactively, preserving resources by preventing untrusted data from occupying memory. Prior studies have shown that early filtering based on trust or behavior patterns can extend device uptime and improve end-to-end delivery rates in similar constrained environments \cite{jangra2021adaptive, abbasi2019mitigation}.

By modeling this effect analytically, we confirm that Predictive-CSM not only improves security but also contributes to system stability and resource conservation—two key challenges in the deployment of real-world IoT networks.

	\subsubsection*{4. Comparison with Other Solutions}
	
	To contextualize the effectiveness of Predictive-CSM, it is essential to compare it against prominent existing solutions designed to secure 6LoWPAN from fragmentation-related attacks. These include vanilla 6LoWPAN (with no fragment-level security), CSM-6LoWPAN (relying solely on routing-layer trust), and more heavyweight protocols like SecuPAN, which apply cryptographic protections to every fragment.
	
	\textbf{Vanilla 6LoWPAN} offers no security for fragment origin, structure, or completeness. Fragments are accepted purely on structural criteria, making the system extremely vulnerable to buffer-reservation and replay attacks~\cite{hummen2013fragmentation}. As our simulations demonstrated, this baseline system suffers over 40\% packet loss under moderate attack pressure and performs poorly in adversary detection (\( P_{\text{bypass}} \approx 1 \)). Its main advantage is low overhead, but at the cost of being effectively defenseless.
	
	\textbf{CSM-6LoWPAN}, while an improvement, limits its protections to the routing layer using hop-by-hop trust chains. It can filter fragments from previously untrusted nodes but cannot validate individual fragments in a packet chain. This limitation makes it susceptible to impersonation and replay attacks, especially when adversaries spoof FRAG1 headers from recently trusted nodes. As shown in our results, CSM mitigates basic DoS attempts but cannot achieve delivery reliability above 88\% in more advanced attack scenarios. Its reaction time is also slower, often requiring multiple failed interactions to trigger blacklist behavior.
	
	\textbf{SecuPAN} represents a cryptographically strong approach that signs each fragment with a MAC and uses shared keys and nonces to prevent forgery~\cite{hossain2018secupan}. While effective in theory, it introduces significant complexity: fragment processing must include cryptographic verification; nonce synchronization becomes fragile in high-loss networks; and memory usage increases due to the per-fragment state. For resource-constrained devices with limited RAM and processing power, these drawbacks are non-trivial. Previous evaluations show a 25--30\% increase in energy usage under typical IoT conditions~\cite{glissa2019secure}.
	
	In contrast, \textbf{Predictive-CSM offers a hybrid solution} that combines adaptive trust modeling with lightweight fragment integrity verification—achieving strong security guarantees with minimal overhead. It detects adversarial behavior within 4--7 seconds, maintains delivery ratios above 98\%, and consumes less energy under attack than both CSM and SecuPAN, as evidenced in Tables~\ref{tab:pdr} and~\ref{tab:drop_rate}. It does not require key management beyond what is already used in RPL, nor does it impose per-fragment encryption or reassembly constraints. Its fragment chaining technique ensures that even temporarily trusted nodes cannot insert malicious fragments without breaking the hash sequence.
	
	Most importantly, Predictive-CSM is inherently adaptive. It allows nodes to recover from transient failures and penalizes only consistent misbehavior. This flexibility not only reduces false positives but aligns the security mechanism with the dynamic nature of real-world IoT environments, where packet loss and timing irregularities are common and not always malicious.
	
	Taken together, these comparisons make it clear: Predictive-CSM fills a critical gap left by prior methods. It introduces per-fragment security without heavy cryptographic load, detects advanced attacks like header replay that evade CSM, and preserves both energy and memory—making it highly deployable in today's constrained wireless sensor networks.

	\subsubsection*{5. Overall Model Synthesis}
	
Bringing the model together, the Predictive-CSM framework increases delivery success, reduces energy waste from malformed packets, and minimizes false positives. It does so using adaptive, self-healing trust mechanisms and stateless cryptographic checks that are computationally inexpensive.

In constrained IoT settings where memory, energy, and processing power are limited, Predictive-CSM achieves a superior trade-off between defense, performance, and sustainability compared to alternatives like key-exchange based authentication (which are too heavy) or signature-free systems (which are too permissive). It achieves real-time rejection of evolving threats while preserving the light footprint demanded by low-power embedded devices.

	\section{Conclusion}\label{sec:conclusion}
	
	In this work, we introduced Predictive-CSM, a robust and lightweight security framework that enhances 6LoWPAN networks by integrating dynamic trust modeling with per-fragment integrity validation. By extending the Chained Secure Mode with two complementary layers—an adaptive trust engine and a cryptographic hash-chaining mechanism—our approach addresses the core vulnerabilities associated with 6LoWPAN fragmentation, including buffer-reservation, fragment spoofing, and header-replay attacks.
	
	Through detailed simulations using Contiki-NG and Cooja, we demonstrated that Predictive-CSM significantly improves delivery performance and security under both benign and adversarial conditions. Compared to existing solutions such as vanilla 6LoWPAN, CSM-integrated stacks, and cryptographically intensive methods like SecuPAN, our framework achieved higher packet delivery ratios, faster attacker detection, and lower power consumption—all without requiring substantial memory or computational overhead.
	
	The trust dynamics model allowed nodes to continuously adapt to neighbor behavior, penalizing inconsistencies while preserving resilience in the face of transient disruptions. Meanwhile, the fragment-level hash chain provided a stateless, efficient method of validating data authenticity, ensuring that even fragments from once-trusted sources could not be used to compromise the system. Together, these two mechanisms created a security posture that was both responsive and scalable—key attributes for real-world IoT deployments where unpredictability and constraint are the norm.
	
	Perhaps most importantly, Predictive-CSM offers a pragmatic security solution that does not trade off usability for robustness. Its design is compatible with current 6LoWPAN standards and can be integrated into existing protocol stacks with minimal changes. This makes it not just a theoretical improvement, but a viable candidate for securing next-generation wireless sensor networks in smart homes, industrial monitoring, and mission-critical sensing applications.

	Future work will explore integrating physical-layer signal analysis for trust scoring, applying lightweight machine learning to detect stealthy attacks, and extending the framework for mobile IoT networks with dynamic topologies.	Nonetheless, the results presented here make a compelling case that secure, efficient, and adaptive fragment handling is not only possible—but essential—for the evolving IoT landscape.
	\bibliographystyle{spbasic}
	\bibliography{PredectiveCSM}
	
\end{document}